**Title:** Vitamin K content of Australian-grown horticultural commodities

**Authors:** Eleanor Dunlop [1,2], Judy Cunningham [2], Paul Adorno [3], Georgios Dabos [3], Stuart K Johnson [4], Lucinda J Black [1,2*]

**Author affiliations:**

[1] Institute for Physical Activity and Nutrition (IPAN), School of Exercise and Nutrition Sciences, Deakin University, Burwood VIC 3125, Australia. e.dunlop@deakin.edu.au; lucinda.black@deakin.edu.au

[2] Curtin School of Population Health, Curtin University, Kent Street, Bentley WA 6102, Australia. eleanor.dunlop@curtin.edu.au; judyc121@gmail.com; lucinda.black@curtin.edu.au

[3] National Measurement Institute, 1/153 Bertie Street, Port Melbourne, VIC 3207, Australia. Paul.Adorno@measurement.gov.au; Georgios.Dabos@measurement.gov.au

[4] Ingredients by Design, Lesmurdie, WA 6076. stuart@ingredientsbydesign.com.au

***Corresponding author:** Lucinda J Black, Institute for Physical Activity and Nutrition (IPAN), School of Exercise and Nutrition Sciences, Deakin University, 221 Burwood Highway, Burwood, VIC 3125, Australia. lucinda.black@deakin.edu.au. Tel.: +61 3 924 45491

**Abstract**

Vitamin K is emerging as a multi-function vitamin that plays a role in bone, brain and vascular health. Vitamin K composition data remain limited globally and Australia has lacked nationally representative data for vitamin K1 (phylloquinone, PK) in horticultural commodities. Primary samples (*n* = 927) of 90 different Australian-grown fruit, vegetable and nut commodities were purchased in three Australian cities. We measured PK in duplicate in 95 composite samples using liquid chromatography with electrospray ionisation-tandem mass spectrometry. The greatest mean concentrations of PK were found in kale (565 μg/100 g), baby spinach (255 μg/100 g) and Brussels sprouts (195 μg/100 g). The data contribute to the global collection of vitamin K food composition data. They add to the evidence that PK concentrations vary markedly between geographic regions, supporting development of region-specific datasets for national food composition databases that do not yet contain data for vitamin K.

## 1. Introduction

Vitamin K is essential for normal blood coagulation and has more recently emerged as an important nutrient for bone, neurological and vascular health (Halder et al., 2019). While other forms of vitamin K (i.e., vitamin K2/menaquinones/MKs) may be found in other foods, vitamin K1 (phylloquinone, PK) is the form of interest in unfermented plant foods (Schurgers & Vermeer, 2000).

PK food composition data remain limited globally, and only a small number of national food composition databases (e.g., US, UK, Denmark, the Netherlands and New Zealand) include data for this important nutrient. The Australian Food Composition Database (Food Standards Australia New Zealand, 2019) does not yet include data for vitamin K. We recently published new analytical data for PK and MK-4 to -10 in nationally representative samples of cheese, yoghurt and meat products purchased in Australia, finding that the K vitamer profile and concentrations varied considerably compared to similar foods analysed in other countries (Dunlop et al., 2022).

PK has been measured in a small number of plant foods sampled from a single Australian city (Palmer et al., 2021); however, more comprehensive data based on national sampling are needed to contribute to the global collection of PK composition data and to allow estimation of population vitamin K intakes in Australia. This is especially important given that dietary vitamin K intake is used as a proxy for vitamin K status in the absence of a single biomarker that can be conveniently measured (Shea & Booth, 2016). Therefore, the aim of this study was to quantify the PK content of Australian-grown horticultural commodities purchased in three Australian cities.

## 2. Methods

### *2.1 Sample purchase and preparation*

Primary samples (*n* = 927) of 90 Australian-grown fruit, vegetable and nut products were purchased in Perth, Sydney and Melbourne during their peak growing season from June 2021 to May 2022 (Table 1). Samples were labelled with a unique sample ID number and couriered overnight from the city of purchase to the National Measurement Institute (NMI) of Australia's Port Melbourne laboratory. Samples were kept chilled from the time of purchase to preparation.

Prior to preparation, samples were photographed to capture appearance (colour, ripeness, condition), variety and packaging/label information on pre-packed items. Samples were prepared (e.g., trimmed, peeled, de-seeded) in the manner that they are usually consumed; peeled and unpeeled samples of cucumber, red-skinned apple and orange sweet potato were requested by the growing industry for analysis of other nutrients and, hence, were included in this study. Equal aliquots of primary samples were homogenised to create a composite sample (*n* = 95) for each commodity, including the aforementioned peeled and unpeeled samples and raw and cooked versions of eggplant and mushrooms. Samples were analysed in the raw form only, except in the case of chestnuts, eggplant and mushrooms. Chestnuts were baked (18 minutes at 200°C) whole in shell and turned halfway through cooking. Cooked and raw samples of eggplant and mushrooms were analysed as there were fewer existing data on nutrient changes upon cooking for these foods in the Australian Food Composition Database (Food Standards Australia New Zealand, 2019). Mushrooms were pan fried without addition of any other ingredient (e.g., oil/fat) until cooked. Whole eggplants were baked (30 minutes at 180°C) and any remaining stem and leaves were removed.

*2.2 Sample analysis*

PK was measured using a verified method that was based on a previously published liquid chromatography electrospray ionisation tandem mass spectrometry (LC-ESI-MS/MS) method (Jäpelt & Jakobsen, 2016; Jensen, Rød, Ložnjak Švarc, Oveland, & Jakobsen, 2022). All procedures were carried out under yellow light. An aliquot of sample weighing 0.5-1.0 g was added to a 50 mL Falcon® tube with a known quantity of labelled internal standard (IS) containing a known concentration of PK-[d7] (Isosciences, Ambler, PA; 100% purity) and 4 mL water (for samples containing fat, water was substituted with 4 ml phosphate buffer (1 M KH2PO4, pH 7.9), 200 µL LecitaseTM Ultra and 200 µL Lipozyme® TL 100L were added and samples were incubated at 37°C overnight in a shaking waterbath). The resulting mixture was vortexed, 25 mL ethanol was added and the mixture was vortexed again. Following addition of 10 mL heptane, the tube was shaken for 30 minutes using a mechanical shaker and centrifuged. A 2 mL heptane layer was taken and evaporated under nitrogen at 40°C. The residue was dissolved in methanol, filtered and transferred to a vial. Separation was performed in a 1290 Infinity II UPLC coupled with a 6490 Triple Quadrupole detector (Agilent Technologies, Santa Clara, USA) and an Ascentis® Express C18 analytical column (2.7 µm, 10 cm x 2.1 mm; Supelco, Bellefonte, PA). The system was set up in electrospray ionisation mode. PK was separated using a gradient elution with 10% methanol in water containing 2.5mM ammonium formate and ethanol containing 2.5mM ammonium formate at a flow rate of 0.5 ml/min and column temperature of 50°C.

Using the aforementioned PK-[d7] IS and a PK IS (Sigma, Purity 99.9%), individual stock standards were prepared for each compound in methanol at concentrations of 50 µg/mL for PK-[d7] and 266.73 µg/mL for PK. Intermediate standards were prepared at 2500.0 ng/mL for PK-[d7] and 2667.33 ng/mL for PK. Six levels of calibration standards were prepared in

methanol with concentrations of 50 ng/mL of PK-[d7] and 6.668, 13.337, 26.673, 53.347, 133.367 or 266.733 ng/mL. PK Calibration curves were obtained by injecting 10 µL of each calibration standard into the system and were used to quantify the concentration of PK in samples.

*2.5 Method verification and comparison with established method*

Three food samples (peeled red-skinned apple, Kent/Jap variety of pumpkin and broccoli) with low, medium and high levels of PK were analysed in the verification study in triplicate. Spike recovery was performed in triplicate at four different spike levels. Duplicated analysis of a control sample (Fapas® 21115, infant formula) was conducted on two different days, one week apart. The same food and control samples, along with high fat samples (olive oil, almond, macadamia and pistachio) were analysed in singlicate using the European standard method with fluorescence detector (FLD) after post column reduction with zinc (European Committee for Standardization, 2003).

*2.4 Quality assurance and data handling*

All samples included in the main study were analysed in duplicate and the average of replicate results for each sample was reported. The relative percent difference (RPD) between replicates was calculated as: (difference between duplicated analyses/average of duplicated analyses) x 100. An RPD of ≤25% was considered acceptable. Fourteen samples were chosen at random for recovery analysis and were spiked with a known concentration of PK. Four samples of a control sample (Fapas® 21115, infant formula) were included in the analysis run. A recovery of 80-120% was considered acceptable. The limit of reporting (LOR) was 0.2 μg/100 g for the majority of matrices; the LOR increased to 0.5 μg/100 g in sweetcorn and passionfruit and to 1.0 μg/100 g in almond and macadamia. Data were aggregated by family.

## 3.0 Results

### *3.1 Method verification and comparison with established method*

The average concentrations from triplicated measurements in red-skinned apple, Kent/Jap pumpkin and broccoli were 0.16, 13.6 and 180.7 μg/100 g, respectively. The respective average spike recovery percentages across four spike levels in the same three foods were 91.5, 96.7 and 101.3%. Values determined for PK in the Fapas® 21115 infant formula control sample were within the reference range (assigned reference value 41.5 μg/100 g, range 26.3-56.6) on both testing days (day 1 replicates = 38.7 and 37.9 μg/100 g; day 2 = 35.8 and 35.1 μg/100 g).

Analysis of the same food and control samples with the published FLD method determined concentrations of 0.37, 12.0 and 178.2 μg/100 g in red-skinned apple, Kent/Jap pumpkin and broccoli, respectively. In high-fat samples, concentrations of 40.2 and 29 μg/100 g were obtained for olive oil and pistachio, respectively. PK was not detected in almond or macadamia using the FLD method (LOR <0.05 μg/100 g). Replicate values of 48.9 and 48.5 g/100 were determined for the Fapas® 21115 infant formula control sample.

### *3.2 PK in horticultural samples*

The mean RPD for PK across all samples was 6.3%. Of 14 spiked samples, the recovery percentage was within 80-120% for 13 (range 79-101%); for one sample (butternut pumpkin), the value (69%) was outside the acceptable range. The recovery percentage for four Fapas® 21115 infant formula control samples ranged from 94 to 109%.

Of 95 samples analysed, 84 contained quantifiable amounts of PK (Table 3). The greatest concentrations were found in kale (565 μg/100 g), baby spinach (255 μg/100 g), Brussels sprouts (195 μg/100 g) and broccoli (186 μg/100 g). Samples that contained the lowest or unquantifiable concentrations of PK included citrus fruits (pink- and yellow-fleshed grapefruit, lemon, lime, mandarin, orange and orange juice), mushrooms (raw and cooked), watermelon, passionfruit, sweetcorn, peeled red-skinned apple and beetroot.

**4.0 Discussion**

This was the first time that PK had been measured in nationally sampled fruit, vegetable and nut products in Australia. We quantified PK in 88% of the horticultural foods sampled. As expected, due to their high chlorophyll content (Booth, 2012), green leafy and cruciferous vegetables contained the greatest concentrations. Kale and baby spinach contained by far the greatest concentrations of foods sampled. Grapefruit, mushroom, orange, orange juice, watermelon, passionfruit and sweetcorn were among samples that did not contain quantifiable concentrations of PK.

Prior to this study, PK had been measured only in horticultural products purchased in one Australian city (Palmer et al., 2021). In that study, two to three individual samples each of 23 different fruit, vegetables, and nuts were purchased from supermarkets in Perth, Western Australia, and analysed individually for PK. Compared to that study, we found similar concentrations in cabbage, cauliflower, celery, green beans, peas, pumpkin, spinach and zucchini, apple and pear. The concentration that we measured in a nationally sampled composite sample of kale was two to five times greater than the range (mean) of 92.5-272.6 (128.5) μg/100 g reported for the three samples analysed in the study by Palmer and colleagues (2021). The concentration of PK in our national composite sample of broccoli was 2-3.5 times that of the range (mean) from the single city study 50.9-92.9 (67.9) μg/100 g. We found slightly lower concentrations of PK in avocado and carrot compared to that study, while some values were difficult to compare without knowing the colour (e.g., capsicum), preparation (peeled/unpeeled) or variety (e.g., lettuce) of samples analysed in that study. In combination, the findings of both studies confirm that PK concentrations can vary markedly within horticultural commodities purchased in Australia and that sampling should be conducted nationally to produce composition data that best represent consumed foods.

Similarly, reported concentrations of PK vary across the globe. The concentration of PK in kale purchased in the Netherlands has been reported to range from 752-881 μg/100 g (Schurgers et al., 2000), which is considerably higher than that found in either Australian study. In contrast, the Danish food composition database presents respective values of 76.9 and 250 μg/100 g for raw 'Danish kale' and 'kale' (National Food Institute, 2019), while the US FoodData Central database (U.S. Department of Agriculture, 2019) contains values ranging from 334 to 418 μg/100 g for raw and cooked kale. These differences may be, in part, attributable to the numerous varieties that may be defined as 'kale'. In the Australian Food Composition Database for example, 'kale' (ID F004780) includes varieties with leaves that may range from a blue-green colour to light green or purple (Food Standards Australia New Zealand, 2019). Such inconsistencies in terminology use represent a challenge globally when comparing food composition data between countries and is another reason why country-specific data are needed (European Food Safety Authority, 2011). PK content is also thought to vary by geographical growing conditions and plant maturity (Ferland & Sadowski, 1992).

Closer to Australia, the highest concentrations of PK in foods measured to date in New Zealand were found in tat soi cabbage (140 μg/100 g) and raw English spinach (110 μg/100 g) (New Zealand Institute for Plant and Food Research & New Zealand Ministry of Health, 2022) at less than half the concentration found in our sample of baby spinach. Corresponding Dutch, Danish and US data for spinach range between 299-429, 340-360 and 381-541 μg/100 g, respectively. While a small number of countries have reasonable or extensive (e.g., the US (U.S. Department of Agriculture, 2019)) PK datasets, the majority of national food composition tables do not yet contain data for vitamin K. It is, therefore, important to keep adding country-specific data to the limited global collection of vitamin K food composition data, especially given the potential extreme variation in content and profile within and between countries.

Intakes of most nutrients were estimated using food consumption data from the 2011-2013 Australian Health Survey (Australian Bureau of Statistics, 2014) and the 2011-2013 Australian Aboriginal and Torres Strait Islander Health Survey (Australian Bureau of Statistics, 2015) and composition data from the Australian Food Composition Database (Food Standards Australia New Zealand, 2019). Due to a lack of suitable data, intakes of vitamin K have not yet been estimated at the national level for the Australian population. Together with the data we produced recently for meat, yoghurt and cheese products (Dunlop et al., 2022), the data from this study contribute to a dataset for the vitamin K content of nationally-sampled foods in Australia. This will allow, for the first time, estimation of PK and MK intakes using nationally-representative food consumption data and monitoring of trends in intake over time. Due to the established variation in vitamin K concentrations, composition data based on national sampling are required for estimating intakes in the Australian population.

Strengths of this study included use of a sensitive and specific verified method, which was based on a previously published and validated method (Jäpelt et al., 2016; Jensen, Ložnjak Švarc, & Jakobsen, 2021; Jensen et al., 2022). Analysis of the FAPAS 21115 infant formula reference material with the LC-ESI-MS/MS method delivered replicated values for PK within the assigned reference range. Quality assurance data were within acceptable ranges, with the exception of the recovery percentage in a spiked sample of butternut pumpkin. As the measured concentration of PK in that sample was low (1.0 μg/100 g), the difference between 69 and 100% recovery represents only a very small (~0.3 μg/100 g) potential difference in actual concentration. The concentrations determined in the reference material and samples of apple, pumpkin and broccoli using the LC-ES-MS/MS and FLD methods were in reasonable agreement. Sampling across three Australian cities that represent the most highly populated regions on both sides of the continent ensured that samples were representative of potential geographical variation. While food composition data is inherently limited in its representation of individual foods, we considered our sampling plan carefully to capture the nutrient content of produce during its peak growing season, when it is most plentiful and most commonly consumed.

A limitation of the study was that our sampling plan was designed for the measurement of a full nutrient profile in a range of horticultural commodities and not specifically for PK measurement. Hence, it did not include some foods (e.g., herbs such as thyme, basil and parsley) that are likely to contain high concentrations of PK (National Food Institute, 2019; U.S. Department of Agriculture, 2019). While it would be useful to determine concentrations in such foods in future studies, herbs are generally consumed in smaller quantities than fruits and vegetables and, therefore, are less likely to contribute substantially to overall PK intake.

It was not feasible to sample produce at multiple time points during the year to capture seasonal variation nor to measure the concentration of PK in individual samples in order to observe variance between them; however, the data produced represent an important advancement in knowledge of the vitamin K content of foods in Australia.

This study has provided data on the vitamin K content of nationally-sampled Australian-grown fruit, vegetable and nut commodities. It adds to a limited but growing collection of data worldwide that demonstrate that the PK content of horticultural commodities varies markedly within and between geographical regions. The data produced in this study contribute to a dataset that will allow the estimation of nationally-representative vitamin K intakes in the Australia for the first time.

**Abbreviations:**

| | |
|---|---|
| AFCD | Australian Food Composition Database |
| FLD | fluorescence detector |
| LC-ESI-MS/MS | liquid chromatography electrospray ionisation tandem mass spectrometry |
| LOR | limit of reporting |
| MK | menaquinone |
| NMI | National Measurement Institute of Australia |
| PK | phylloquinone |
| RPD | relative percent difference |

**Financial support:** This study was supported by the Western Australian Department of Health Future Health Research and Innovation (FHRI) Fund. The sampling phase of this work was supported by Hort Innovation (ST19036). Authors worked with Hort Innovation to develop the scope for this project and to identify commodities of priority for inclusion. Authors wrote the sampling plan, which was approved by Hort Innovation. Hort Innovation had no role in the conduct of the study, preparation and interpretation of the study's findings or writing of this article. ED is supported by a Multiple Sclerosis Australia Postdoctoral Fellowship. LJB is supported by Multiple Sclerosis Western Australia (MSWA), and a Multiple Sclerosis Australia Postdoctoral Fellowship.

**Author contributions: Eleanor Dunlop:** Conceptualization, Funding acquisition, Methodology, Data curation, Project administration, Writing – original draft **Judy Cunningham:** Conceptualization, Funding acquisition, Methodology, Data curation, Writing – review and editing **Paul Adorno:** Methodology, Funding acquisition, Resources, Writing – review and editing **Georgios Dabos:** Methodology, Investigation, Writing – original draft **Stuart Johnson:** Funding acquisition, Project administration, Writing – review and editing **Lucinda J Black:** Conceptualization, Funding acquisition, Methodology, Supervision, Writing – review and editing

**Table 1** Characteristics of Australian-grown horticultural samples purchased in Sydney, Melbourne and Perth for analysis of phylloquinone content

| Sample | Botanical name, alternate common names and varietal names | Primary sample purchases | Total number of pieces purchased | Total weight of purchases (kg) |
|---|---|---|---|---|
| Almond, Raw, with skin | *Prunus dulcis* | 10 | 11 packets | 4.91 |
| Apple, Golden Delicious variety, Unpeeled | *Malus domestica* - Golden Delicious | 9 | 52 | 6.68 |
| Apple, Granny Smith variety, Unpeeled | *Malus domestica* - Granny Smith | 10 | 53 | 6.53 |
| Apple - Pink Lady variety, Peeled | *Malus domestica* - Cripps Pink | 10 | 20.5 | 3.03 |
| Apple - Pink Lady variety, Unpeeled | *Malus domestica* - Cripps Pink | 10 | 57 | 3.54 |
| Apricot, Dried | *Prunus armeniaca.* | 9 | 9 packets | 1.90 |
| Apricot, Fresh | *Prunus armeniaca*. | 10 | 111 | 6.39 |
| Avocado, Hass variety | *Persea americana.* | 8 | 28 | 6.35 |
| Avocado, Shepherd variety | *Persea americana.* | 10 | 41 | 5.51 |
| Banana, Cavendish variety | *Musa acuminata* | 10 | 38 | 6.00 |
| Banana, Lady Finger variety | *Musa acuminata*. Sugar banana. | 10 | 52 | 4.75 |
| Bean, Green | *Phaseolus vulgaris*. French beans, string beans. | 10 | 12 packets | 5.23 |
| Beetroot, Root only | *Beta vulgaris*. Red-coloured root only. | 10 | 46 | 11.91 |
| Blackberry | *Rubus* species | 10 | 30 punnets | 3.38 |
| Blueberry | Genus *Vaccinium*. | 10 | 30 punnets | 4.17 |
| Bok choy | *Brassica rapa subsp. Chinensis.* Baby bok choy, Buk choy, Pak choy | 8 | 27 | 3.90 |
| Broccoli | *Brassica oleracea var italica.* Dark green in colour. | 8 | 16 heads | 5.29 |
| Brussels sprouts | *Brassica oleracea var. gemmifera* | 12 | 202 | 4.76 |
| Cabbage, Red/Purple | *Brassica oleracea var capitata* | 8 | 2 x whole, 6 x halves | 6.08 |
| Cabbage, White | *Brassica oleracea var capitata*. Common cabbage. | 11 | 2 x whole, 8 x halves, 1 x quarter | 9.22 |
| Capsicum, Green | *Capsicum annuum*. Bell pepper | 10 | 20 | 4.04 |
| Capsicum, Red | *Capsicum annuum*. Bell pepper | 10 | 15 | 3.79 |
| Capsicum, Yellow | *Capsicum annuum*. Bell pepper | 10 | 15 | 3.29 |
| Carrot, Mature | *Daucus carota*. Orange coloured varieties. | 10 | 68 | 7.53 |
| Cauliflower, White | *Brassica oleracea var. botrytis* | 8 | 6 x whole, 2 halves | 6.66 |
| Celery | *Apium graveolens* | 10 | 9 bunches | 10.34 |
| Cherry | Genus *Prunus* | 10 | 14 boxes | 6.41 |
| Chestnut, Roasted, Peeled | *Castanea sativa* | 10 | 34-56 per purchase | 5.56 |
| Chilli, Red Thai-style | *Capsicum annuum*. Long red chilli | 10 | 138 | 2.65 |
| Cucumber, Lebanese variety, Peeled | *Cucumus sativus*. Dark green skin. | 10 | 31 | 4.30 |
| Cucumber, Lebanese variety, Unpeeled | *Cucumus sativus*. White flesh. | 10 | 38 | 4.99 |
| Currant (dried grape) | *Vitis vinifera* | 7 | 9 packets | 3.23 |
| Custard apple | *Annona squamosa x Annona cherimola.* African Pride and Pinks Mammoth are common varieties. | 9 | 28 | 10.85 |
| Eggplant | *Solanum melangena*. Aubergine. | 10 | 21 | 8.28 |
| Grapefruit, Pink fleshed | *Citrus x Paradisi.* Ruby (ruby blush) grapefruit. | 9 | 39 | 11.66 |
| Grapefruit, Yellow fleshed | *Citrus x Paradisi.* Marsh is a common variety. | 7 | 28 | 10.10 |
| Grape, Green | *Vitis vinifera* | 10 | 10 bags | 3.87 |
| Grape, Red | *Vitis vinifera* | 10 | 10 bags | 7.79 |
| Honeydew melon | *Cucumis melo* | 10 | 10 | 7.34 |
| Kale | *Brassica oleracea var. sabellica.* Tuscan kale, curly kale, baby kale. | 8 | 8 bunches | 2.98 |
| Leek | *Allium porrum* | 10 | 21 | 5.30 |
| Lemon, flesh | *Citrus limon* | 8 | 50 | 9.26 |
| Lettuce, Cos variety | *Lactuca sativa L. var. longifolia.* Romaine lettuce, green coloured. | 10 | 19 | 3.93 |
| Lime, flesh | *Citrus aurantifolia*. Tahitian lime | 8 | 74 | 7.62 |
| Lychee, Fresh | *Litchi chinensis* | 10 | 221 | 4.17 |
| Macadamia | *Macadamia integrifolia*. Queensland nut. | 10 | 19 packets | 4.77 |
| Mandarin | *Citrus reticulata*. Purchased varieties include Imperial, Murcott, Afourer | 19 | 117 | 8.95 |
| Mango, Kensington Pride | *Mangifera indica*. Bowen mango. | 9 | 34 | 8.39 |
| Mushroom, Common, Cooked | *Agaricus bisporus* | 10 | 276 | 6.33 |
| Mushroom, Common, Raw | *Agaricus bisporus* | 10 | 164 | 4.16 |
| Nashi pear | *Pyrus pyrifolia*. Nashi, Asian pear. | 10 | 32 | 6.53 |
| Nectarine, White fleshed | *Prunus persica var. nucipersica* | 10 | 64 | 6.26 |
| Nectarine, Yellow fleshed | *Prunus persica var. nucipersica* | 10 | 63 | 5.84 |
| Olive oil | *Olea europaea* | 10 | 10 bottles | 8.20 |
| Olives, Preserved | *Olea europaea* | 10 | 22 jars | 8.61 |
| Onion, Brown skin | *Allium cepa* | 10 | 50 | 5.18 |
| Onion, Red skin | *Allium cepa* | 10 | 45 | 6.14 |
| Onion, White skin | *Allium cepa* | 12 | 45 | 5.93 |
| Orange | *Citrus sinensis*. Mixture of Navel and Valencia types, likely to be of multiple varieties. Red-fleshed oranges not included. | 10 | 49 | 14.24 |
| Orange juice, Fresh, Australian | | 10 | 11 bottles | 11.66* |
| Papaya, Red fleshed | *Carica papaya* | 10 | 8 x whole, 2 x halves | 9.13 |
| Papaya, Yellow fleshed | *Carica papaya*. Pawpaw. | 10 | 6 x whole, 4 x halves | 5.56 |
| Parsnip | *Pastinaca sativa* | 10 | 55 | 6.66 |
| Passionfruit | *Passiflora edulis* | 18 | 195 | 7.46 |
| Peach, White fleshed | *Prunus persica* | 10 | 58 | 5.97 |
| Peach, Yellow fleshed | *Prunus persica* | 10 | 68 | 6.72 |
| Pear, Brown skinned | Genus *Pyrus* | 10 | 28 | 5.80 |
| Pear, Green skinned | Genus *Pyrus*. Packham pear, Bartlett pear. | 10 | 31 | 6.90 |
| Pea, Green, In pod | *Pisum sativum* | 8 | 9 packets | 6.74 |
| Persimmon | *Diospyros kaki*, may include sweet or astringent varieties | 6 | 28 | 4.61 |
| Pineapple | *Ananus comosus*. | 10 | 10 | 7.69 |
| Pistachio | *Pistacia vera* | 8 | 16 packets | 1.35 |
| Plum | *Prunus domestica*. | 10 | 70 | 6.00 |
| Potato, Unpeeled, Raw | *Solanum tuberosum*. Includes potatoes sold as red-skinned, white-skinned or kipfler. | 10 | 84 | 10.55 |
| Prune | *Prunus domestica* | 9 | 11 packets | 4.78 |
| Pumpkin, Mixed varieties, including Kent, Butternut, Golden Nugget and Japanese-style) | *Cucurbita pepo* | 10 | 2 x whole, 3 x halves, 6 x quarters, one sixth | 14.86 |
| Pumpkin, Butternut variety | *Cucurbita pepo* | 10 | 9 x whole, 1 x half | 5.91 |
| Pumpkin, Kent/Jap variety | *Cucurbita pepo* | 10 | 1 x whole, 1 x half, 8 x quarters | 7.15 |
| Raisin (dried grape) | *Vitis vinifera* | 7 | 11 packets | 4.26 |
| Raspberry | *Rubus* species | 10 | 30 punnets | 3.40 |
| Rockmelon | *Cucumis melo*. Canteloupe. | 10 | 10 | 6.55 |
| Silverbeet | *Beta vulgaris*. Chard. Dark green leaves. | 10 | 13 bunches | 8.20 |
| Snow pea | *Pisum sativum* | 10 | 22 packets | 4.84 |
| Spinach, Baby | *Spinacea oleracea* | 10 | 17 bags | 4.06 |
| Strawberry | Genus *Fragaria*. | 9 | 261 | 4.13 |
| Sultana | *Vitis vinifera* | 10 | 10 packets | 4.86 |
| Sun muscat | *Vitis vinifera* | 10 | 12 packets | 3.97 |
| Sweet Potato, Orange fleshed-Peeled | *Ipomoea batatas*. Kumara. | 19 | 39 | 7.73 |
| Sweet Potato, Orange fleshed-Unpeeled | *Ipomoea batatas*. Kumara. | 10 | 4 x whole, 6 x halves | 1.64 |
| Sweet Potato, Purple fleshed | *Ipomoea batatas.* | 7 | 23 | 11.41 |
| Sweet Potato, White fleshed | *Ipomoea batatas*. Generally have brown skin | 9 | 26 | 8.78 |
| Sweetcorn | *Zea mays* convar. *saccharata* var. *rugosa* | 10 | 33 cobs | 10.30 |
| Watermelon | *Citrulis lannatus* | 10 | 10 x halves | 9.73 |
| Zucchini | *Cucurbita pepo*. Courgette. | 10 | 37 | 7.14 |

*unit of measurement is L

**Table 2** Phylloquinone content in the edible portion of Australian-grown horticultural commodities purchased in Sydney, Melbourne and Perth, aggregated by family

| Food type | Individual samples analysed | Phylloquinone (µg/100 g) Mean | SD |
|---|---|---|---|
| Apples and pears | Apple - red skinned, peeled; Apple - Golden Delicious, unpeeled; Apple - red skinned, unpeeled; Apple - Granny Smith, unpeeled; Pear, brown skinned; Pear, green skinned; Pear, nashi | 2.4 | 1.4 |
| Berries | Blackberry; Blueberry; Raspberry; Strawberry | 8.6 | 4.8 |
| Citrus fruit | Grapefruit, pink-fleshed; Grapefruit, yellow-fleshed; Lemon flesh; Lime flesh; Mandarin; Orange; Orange Juice, fresh, Australian | 0.1 | 0.1 |
| Vine fruit | Currant (dried grape); Grape, green; Grape, red; Honeydew melon; Rockmelon; Watermelon; Raisin (dried grape); Sultana; Sun muscat | 11.4 | 11.4 |
| Tropical fruit | Banana, Cavendish variety; Banana, Lady Finger variety; Custard apple; Lychee, fresh; Mango, Kensington Pride; Papaya, red-fleshed; Papaya, yellow-fleshed; Passionfruit; Persimmon; Pineapple | 1.4 | 2.5 |
| Stone fruit | Apricot, dried; Apricot, fresh; Cherry; Nectarine, white-fleshed; Nectarine, yellow-fleshed; Peach, white-fleshed; Peach, yellow-fleshed; Plum; Prune | 11.0 | 19.8 |
| Brassica vegetables | Bok choy; Broccoli; Brussels Sprouts; Cabbage, red/purple; Cabbage, white; Cauliflower, white; Kale | 160.3 | 191.7 |
| Roots and tubers | Beetroot, root only; Carrot, mature; Celery; Onion, brown-skinned; Onion, red-skinned; Onion, white-skinned; Parsnip; Potato, unpeeled, raw; Sweet potato, orange-fleshed, peeled; Sweet potato, orange-fleshed, unpeeled; Sweet potato, purple-fleshed; Sweet potato, white-fleshed | 3.1 | 5.9 |
| Legumes and corn | Bean, green; Pea, green; Snow pea; Sweetcorn | 24.0 | 18.3 |
| Leaf and stem green vegetables | Leek; Lettuce, Cos variety; Silverbeet; Spinach, baby | 102.7 | 107.9 |
| Fruiting vegetables | Avocado (Hass); Avocado (Shepherd); Capsicum, Green; Capsicum, red; Capsicum, yellow; Chilli, red Thai-style; Cucumber, peeled; Cucumber, unpeeled; Eggplant, oven-roasted; Eggplant, raw; Olive oil; Olives, preserved; Pumpkin, mixed varieties; Pumpkin, Butternut variety; Pumpkin, Kent/Japanese-style variety; Zucchini | 12.1 | 10.6 |
| Nuts | Almond, raw, with skin; Chestnut, roasted, peeled; Macadamia; Pistachio | 6.5 | 12.7 |
| Funghi | Mushroom, common, raw; Mushroom, common, cooked | 0 | 0 |

All values are expressed on a fresh weight basis. Composite samples of individual commodities, comprising equal aliquots of samples purchased in Sydney, Melbourne and Perth, were analysed. Concentrations measured in composite samples are an average of duplicated analyses.

Limit of quantification range: 0.2 µg/100 g for all samples except: sweetcorn and passionfruit (0.5 µg/100 g) and almond and macadamia (1.0 µg/100 g)